# Leaky-Wave Antenna for Non-Invasive Continuous Glucose Monitoring

Huan Wang, Han Xiao, Yi Zhang, Zhi Hua Ren, *Member, IEEE,* Yang Gao, *Member, IEEE*

***Abstract*—Non-invasive continuous glucose monitoring requires sensing techniques capable of delivering stable, accurate, and real-time quantification of glucose-induced dielectric variations. This paper presents a microwave-based methodology that employs a leaky-wave antenna (LWA) to launch localized traveling electromagnetic waves and estimate glucose concentration through phase measurements at a fixed operating frequency. The LWA architecture enhances electromagnetic coupling with the material under test through controlled energy leakage, enabling strong linearity and high phase sensitivity within the physiological permittivity range (relative permittivity 68–70). In vitro measurements using fetal bovine serum–glucose samples demonstrate a total phase shift of 8.51° across glucose concentrations from 50 to 255 mg/dL. Preliminary in vivo experiments conducted on two human subjects show strong correlations between measured $S_{11}$ phase and reference glucose levels (r = 0.939 and r = 0.906), confirming the potential of the LWA-based sensing mechanism to track dynamic glycemic variations in a proof-of-concept setting. These results highlight the potential of the proposed approach for practical, continuous, and non-invasive glucose monitoring.**

***Index Terms*—Non-invasive sensing, continuous glucose monitoring (CGM), leaky-wave antenna (LWA), electromagnetic sensor, dielectric property measurement.**

## I. INTRODUCTION

Glucose monitoring is critical for effective management of diabetes and for advancing personalized healthcare strategies [1]. For individuals with diabetes, regular and accurate glucose monitoring is essential to prevent acute complications [2], including kidney disease, stroke, and cardiovascular diseases. Additionally, continuous monitoring of blood glucose levels during pregnancy plays a vital role in managing glycemic control, helping mitigate the risks of gestational diabetes [3] and reducing complications such as miscarriage, polyhydramnios, and macrosomia. As a result, the development of effective blood glucose monitoring sensors is of significant importance.

Blood glucose monitoring techniques can be broadly categorized into invasive and non-invasive methods, based on whether they require direct access to blood samples [4]. Invasive methods primarily include finger-prick testing [5] and subcutaneous sensor monitoring [6]. These approaches are well established for their accuracy, although interest in non-invasive alternatives continues to grow due to the desire for more comfortable and seamless long-term monitoring.

Non-invasive glucose monitoring methods [4] are further classified into optical, electrochemical, and microwave techniques. Optical methods [7] offer versatile optical probing mechanisms for biochemical assessment. Electrochemical approaches [8], [9] enable selective molecular detection through engineered sensing materials. Alongside these, microwave glucose sensors [10], [11] have emerged as a promising direction due to the strong relationship between glucose concentration in biological fluids and corresponding dielectric properties [12]. By sensing variations in the permittivity of tissue or blood, microwave systems hold significant potential for developing non-contact and continuous monitoring capabilities.

Microwave sensing techniques [13] are commonly divided into resonant [14], [15] and non-resonant [16], [17] categories. Resonant sensors, such as split-ring resonators (SRRs) [18] and complementary split-ring resonators (CSRRs) [19], [20], achieve high sensitivity through strong confinement of electromagnetic energy at discrete operating frequencies. These resonant structures provide sharp spectral features that can be leveraged to detect small dielectric variations and have been widely adopted in glucose-related studies [15].

This paper presents a non-resonant microwave sensing approach based on a leaky-wave antenna (LWA) architecture. Instead of relying on resonance phenomena, the LWA supports a traveling-wave mode that allows electromagnetic energy to propagate progressively along its aperture, providing a complementary mechanism for detecting glucose-induced dielectric variations through phase measurements at a fixed operating frequency.

The sensing principle is implemented by monitoring the phase variation of the reflection coefficient ($S_{11}$), which

This work was supported in part by National Natural Science Foundation of China (52307256), the Fundamental Research Funds for the Central Universities (XJSJ23009), Proof of Concept Foundation of Xidian University Hangzhou Institute of Technology (GNYZ2023YL0404).
(Corresponding authors: Yang Gao, e-mail: gaoyang.ziint@gmail.com; Zhi Hua Ren, e-mail: renzhh@shanghaitech.edu.cn)

Huan Wang is with Hangzhou Institute of Technology, Xidian University, 311200 Hangzhou, China.

Han Xiao is with Hangzhou Institute of Technology, Xidian University, 311200 Hangzhou, China.

Yi Zhang is with the Department of Radiology, The Second Affiliated Hospital, Zhejiang University School of Medicine, Hangzhou, China

Zhi Hua Ren is with the School of Biomedical Engineering & State Key Laboratory of Advanced Medical Materials and Devices, ShanghaiTech University, Shanghai, 201210, China, and he is also with the Shanghai Clinical Research and Trial Center, Shanghai, 200231, China

Yang Gao is with the Sir Peter Mansfield Imaging Centre, School of Physics and Astronomy, University of Nottingham, Nottingham, UK, and he is also with the Hangzhou Institute of Technology, Xidian University, 311200 Hangzhou, China correlates directly with permittivity changes associated with glucose concentration in skin tissue. The proposed LWA sensor is systematically validated through full-wave electromagnetic simulations, in-vitro measurements using serum–glucose samples, and preliminary in-vivo human trials. Experimental results demonstrate a strong linear correlation between the measured phase response and reference glucose concentrations, confirming the feasibility of the non-resonant, phase-based approach for robust and continuous glucose monitoring applications.

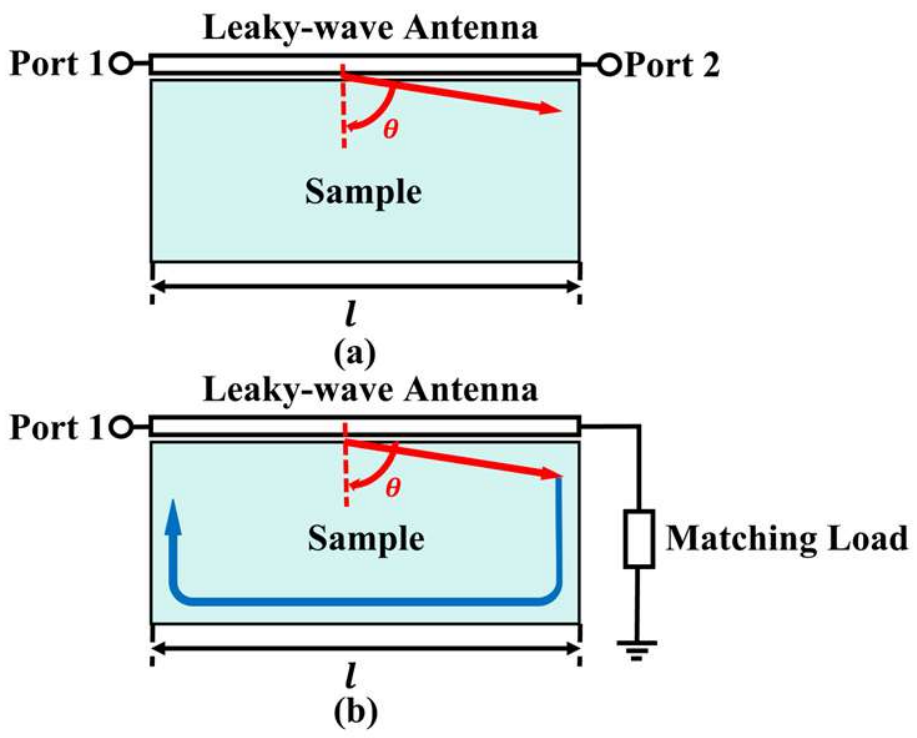

**Fig. 1.** Schematic of leaky-wave antenna glucose sensing modes:(a) two-port transmission measuring phase shift; (b) one-port reflection measuring.

## II. THEORY AND METHOD

### *A. Working principle*

To understand the operating principle of the proposed leaky-wave antenna (LWA) sensor, it is instructive to first examine the characteristics of its foundational structure—the traveling-wave antenna. Traveling-wave antennas radiate by supporting a wave that propagates along a guiding structure, with radiation occurring continuously along the path [21]. In conventional microstrip implementations, this wave interacts weakly with the surrounding medium due to limited field leakage.

To address this limitation, LWAs incorporate periodic or discontinuous perturbations that enable controlled leakage of electromagnetic energy into adjacent media, thereby enhancing coupling with the tissue. The resulting radiation is progressive along the guiding structure and exhibits a well-defined radiation angle $\theta$, which is directly related to the phase constant $\beta$ in the antenna via [22]

$$\theta = sin^{-1}(\frac{\beta}{k}) \tag{1}$$

where $k$ is the wave-number in the medium under test (MUT).

This controlled field leakage promotes stronger field–tissue interaction, making LWAs particularly well-suited for broadband, non-resonant dielectric sensing applications.

We examine two possible configurations for the proposed LWA sensor, as shown in Fig. 1. In the first configuration, Fig. 1(a), the antenna operates in a two-port transmission mode. The phase of the transmission coefficient $\Delta\varphi_{S_{21}}$ reflects the phase evolution along wave propagation path inside the MUT:

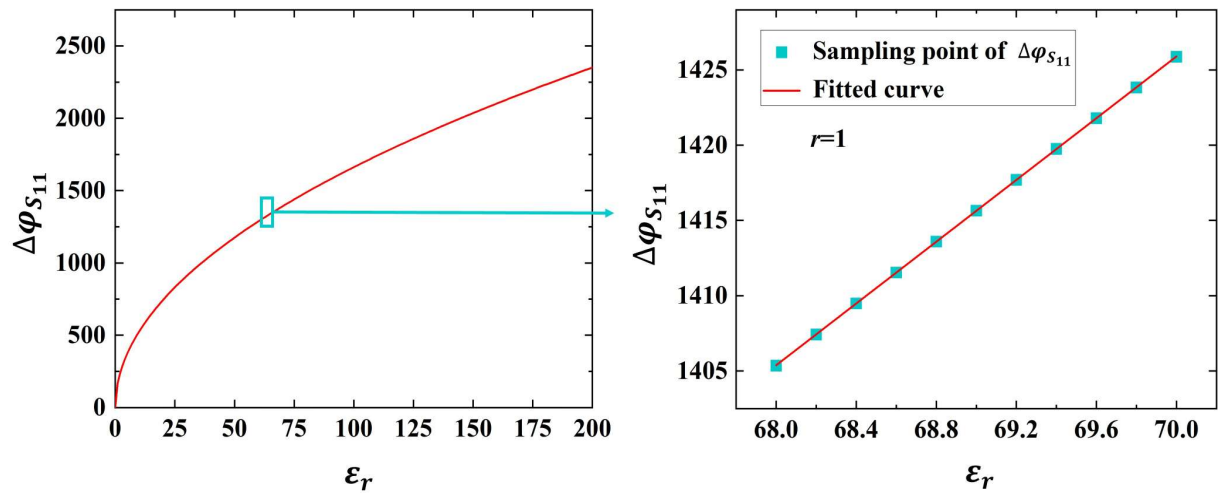

**Fig. 2.** (a) Simulated phase shift $\varphi_{S11}$ of the $S_{11}$ parameter as a function of the material permittivity $\varepsilon_r$ at 2.63 GHz, following the theoretical relationship $\Delta\varphi_{S1} = 2 \cdot k_0 \cdot l \cdot \sqrt{\varepsilon_r}$. (b) Zoomed-in view of the physiological permittivity range $\varepsilon_r$ =68-70, showing discrete sampling points and the nearly linear phase response used for local linear fitting (*r*=1).

$$\Delta\varphi_{S_{21}} = \int_0^l k\,(l)dl \tag{2}$$

Here, $l$ is the effective length in propagation, and the wave-number in MUT $k$ is related to the medium's dielectric constant $\varepsilon_r$ by (simplified analysis for the lossless case):

$$k = k_0\sqrt{\varepsilon_r} \tag{3}$$

Alternatively, a matching load – such as 50 Ω terminator – can be connected to one port of the LWA to suppress wave reflection and establish traveling-wave operation, as shown in Fig.1(b). By measuring the phase shift of the reflection coefficient $\Delta\varphi_{S_{11}}$, the propagation characteristics within the MUT can also be extracted:

$$\Delta\varphi_{S_{11}} = 2\int_0^l k\,(l)dl \tag{4}$$

It is noteworthy that the total phase evolution observed in $\Delta\varphi_{S11}$ is theoretically expected to be twice that of the transmission measurement $\Delta\varphi_{S21}$. Accordingly, the second configuration exhibits a phase sensitivity that is twice as high as that of the first configuration. As illustrated in Fig. 2(a), the simulated $S_{11}$ phase response of the leaky-wave antenna exhibits a nonlinear relationship with the square root of the dielectric constant ($\Delta\varphi_{S_{11}} \propto \sqrt{\varepsilon_r}$). However, within the physiologically relevant range for blood glucose measurement, the relative permittivity of biological tissue varies only slightly (from $\varepsilon_r$=68-70) [23], resulting in a nearly linear segment of the curve. This is further demonstrated in Fig. 2(b), where the discrete phase response data points within this narrow interval show an excellent intrinsic linear correlation (*r*=1) inherent to noise-free deterministic simulations. Therefore, for practical purposes, the relationship between phase shift and relative permittivity can be accurately approximated by a linear model in this range, which greatly simplifies both data processing and quantitative analysis of glucose-induced dielectric changes.

TABLE I
DESIGN PARAMETERS AND THEIR VALUES OF THE LWA

| $L$ | $l_{s1}$ | $l_{s2}$ | $W$ | $w_m$ | $w_s$ |
|---|---|---|---|---|---|
| 4.40mm | 3.70mm | 4.50mm | 5.09mm | 1.00mm | 0.35mm |
| $h_1$ | $h_2$ | $\varepsilon_{h1}$ | $\varepsilon_{h2}$ | $tan\delta_{h_1}$ | |
| 0.80mm | 0.11mm | 4.30 | 1.04 | 0.02 | |

TABLE II
DESIGN PARAMETERS AND THEIR VALUES OF THE CSRR

| $L$ | $W$ | $h$ | $\varepsilon_r$ | $tan\delta$ | $W_m$ |
|---|---|---|---|---|---|
| 68.00mm | 20.00mm | 0.80mm | 4.30 | 0.02 | 1.50mm |
| $a$ | $b$ | $C_w$ | $g$ | $s$ | $t$ |
| 7.60mm | 7.60mm | 12.00mm | 0.40mm | 0.40mm | 0.40mm |

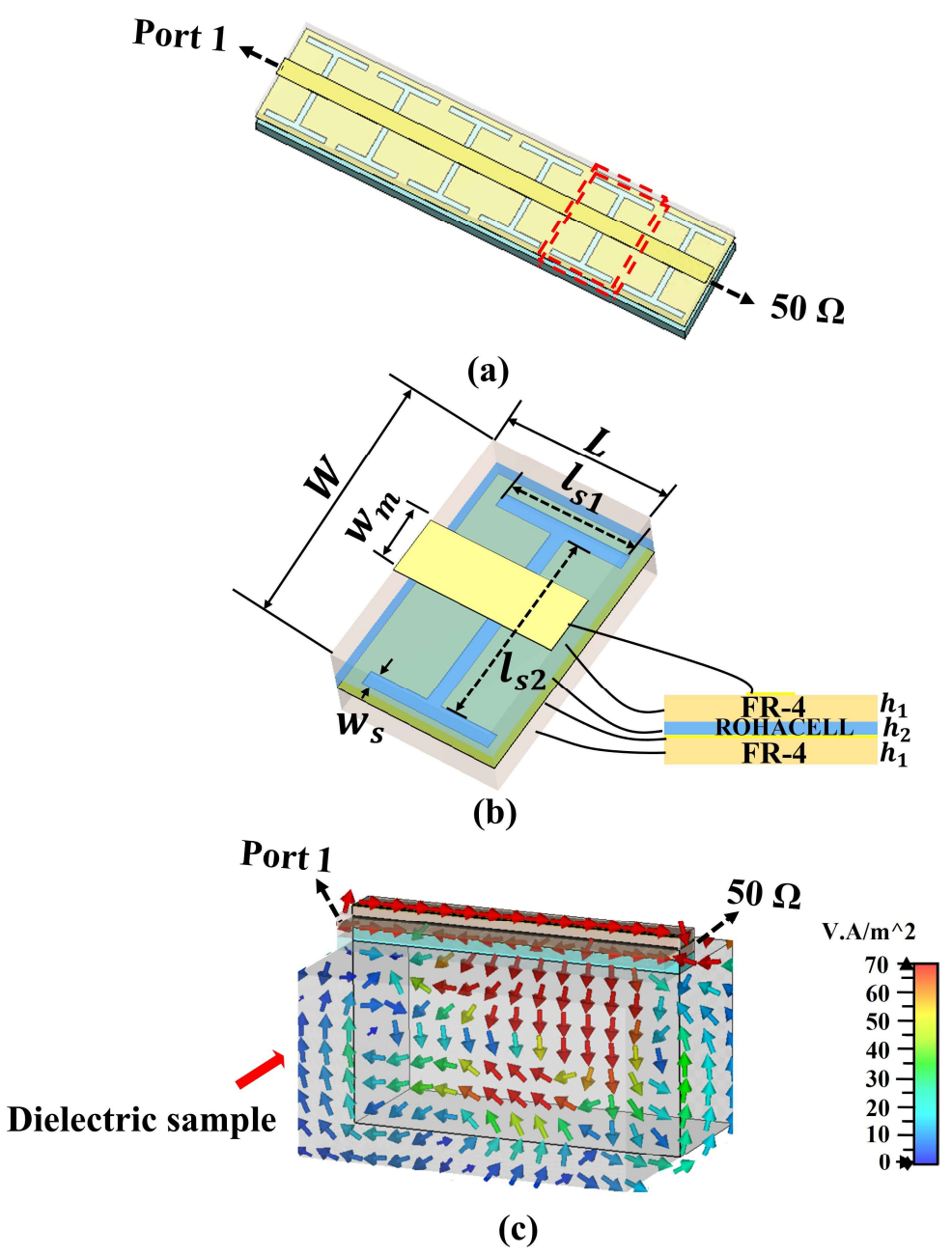

**Fig. 3.** (a) The overall structure of the LWA. (b) The unit cell of the LWA. (c) Power-flow streamlines of the LWA at 2.63 GHz from CST Microwave Studio 2021, illustrating progressive leakage of electromagnetic energy along the slot into the dielectric sample.

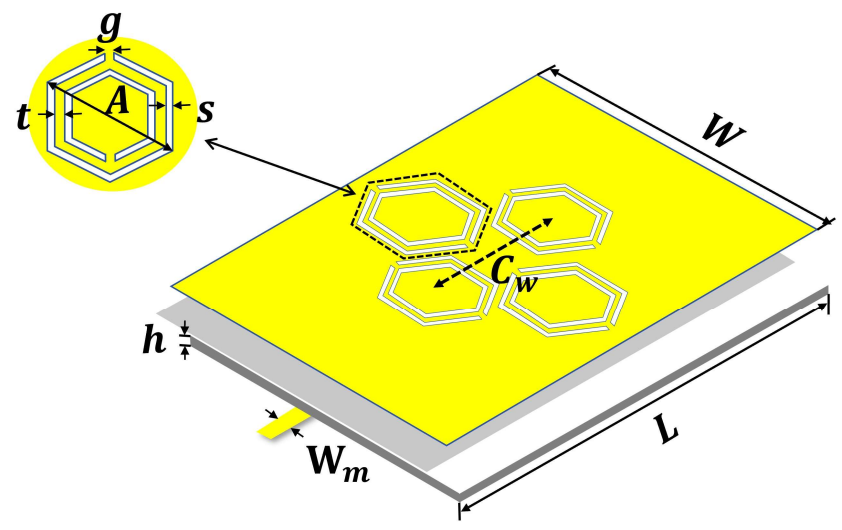

**Fig. 4.** The overall structure of the CSRR. Schematic of the CSRR sensor structure. Four hexagonal complementary split-ring resonators (CSRRs) are etched in the ground plane of an FR-4 substrate (thickness $h$, relative permittivity $\varepsilon_r$), with a microstrip feed line of width $W_m$ on top. The inset shows the geometric parameters: $A$ (hexagon side), $g$ (split gap), $s$ (spacing), and $t$ (ring width). $C_w$ indicates the center-to-center distance between adjacent CSRRs.

The physical sensing mechanism can be summarized as follows:

1. Variations in blood glucose concentration modify the complex dielectric properties (permittivity and conductivity) of blood and interstitial fluid (ISF).
2. The LWA launches a guided traveling electromagnetic wave along the sensor–tissue interface.
3. The phase velocity of this wave depends on the dielectric properties of the surrounding medium.
4. The measured phase shift of the reflection coefficient represents the accumulated phase delay along the wave's round-trip propagation path.
5. Thus, by measuring the phase of $S_{11}$ at a fixed frequency, glucose-induced dielectric variations are directly quantified through their effect on propagation delay.

### B. Sensor design

While the reflection-based measurement approach enhances sensitivity, its performance remains highly dependent on the sensor's structural and material properties. Therefore, we designed a leaky-wave antenna (LWA) sensor with optimized field leakage and dielectric coupling, as detailed below.

Fig. 3(a) illustrates the overall structure of the proposed leaky wave antenna. The proposed sensor adopts a multilayer microstrip configuration to balance electromagnetic leakage, structural compactness, and tissue conformity. From top to bottom, the antenna consists of a microstrip line for excitation, a dielectric substrate, a foam spacer layer, a patterned ground plane with radiation slots and a matching dielectric layer.

Moreover, the sensor's slot design underwent several iterations, ultimately resulting in the final configuration of six 'I'-shaped slots. This design was chosen to achieve optimal power leakage to the sample and to optimize the sensor's dimensions for complete coverage of the finger. The ROHACELL 31HF foam layer, placed between the two PCB layers, plays a crucial role in tuning the sensor's operating frequency. Additionally, using CST Microwave Studio 2021, we performed a power-flow analysis of the LWA at 2.63 GHz. As depicted in Fig. 3(c), the resulting power-flow vectors demonstrate that the majority of electromagnetic energy leaks progressively along the slot into the dielectric sample region, validating our slot-geometry optimization for maximal absorption. The working frequency of approximately 2.6 GHz was chosen to achieve a trade-off between adequate subdermal penetration—given the inverse dependence of penetration depth on frequency—and enhanced sensitivity to glucose-induced permittivity changes [14], while also accommodating the size constraints of wearable sensor integration.

### C. Numerical simulations

A series of detailed simulations were carried out in CST Microwave Studio 2021 to investigate the influence of slot geometry on sensor performance. Specifically, the effects of slot length and slot width on the energy absorption efficiency and phase sensitivity were systematically analyzed.

Fig. 3(b) presents a perspective view of a single unit cell in the leaky-wave antenna structure. The top copper layer forms a stripline, while the bottom copper layer functions as a ground plane with six etched 'I'-shaped slots. Detailed parametric design parameters for the LWA are provided in Table II. These conductive layers are supported by two 0.8 mm thick single-layer FR-4 substrates ($\varepsilon_r$=4.3, $tan\delta$ = 0.02), separated by a 0.11 mm thick ROHACELL 31HF foam spacer with a dielectric constant of 1.04. The LWA antenna is driven by a 50 Ω coaxial cable, which is soldered to the microstrip line at one end. To minimize wave reflections and ensure traveling-wave excitation, the other end is terminated with a matched 50 Ω resistive load.

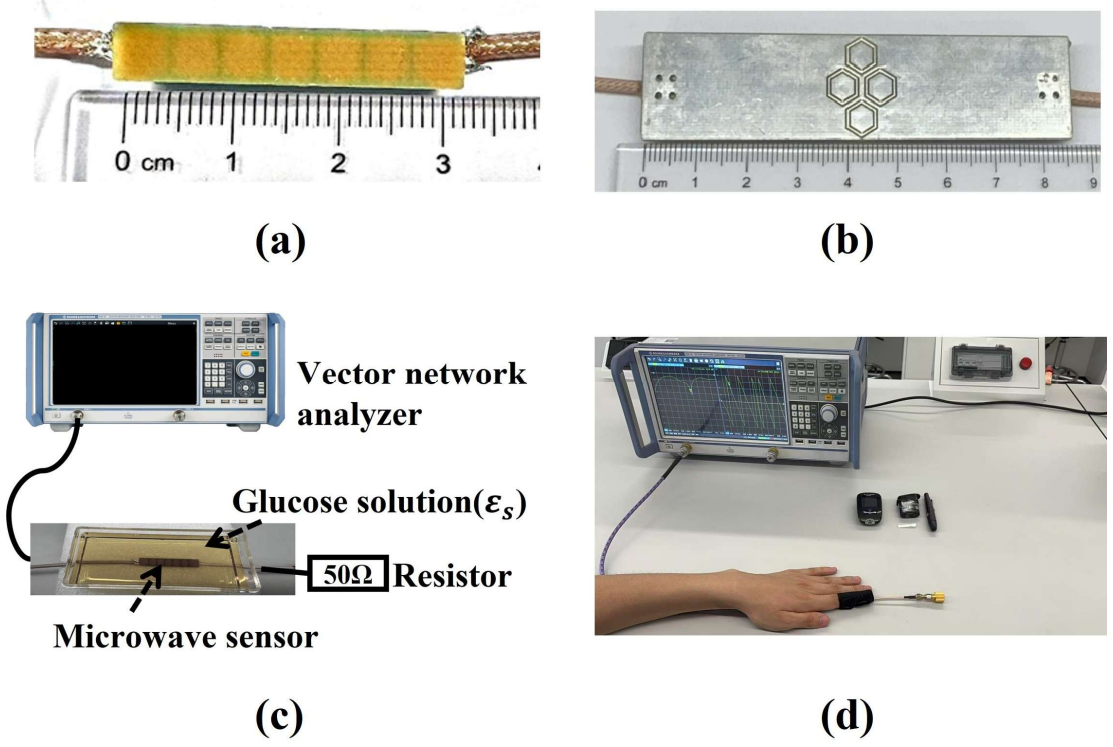

**Fig. 5.** (a) Fabricated prototype of the LWA sensor. (b) Fabricated prototype of the CSRR sensor. (c) Experimental setup of the glucose monitoring system using the fabricated microwave sensor. (d) Experimental setup for in-vivo glucose measurements in humans.

By optimizing these parameters, the optimal slot configuration for both the CSRR and LWA sensors was established to maximize absorbed power and enhance sensitivity within the physiological permittivity range. These simulation results highlight the critical relationship between slot structure, sensor sensitivity, and energy absorption in the dielectric sample, providing valuable guidance for the subsequent experimental design.

To highlight the unique advantage of the proposed LWA sensor over conventional resonant design, a classic complementary split-ring resonator (CSRR) structure—widely adopted for high-sensitivity dielectric sensing [24]—was also implemented for comparison. This sensor (Fig.4) consists of four identical hexagonal resonators arranged in a honeycomb structure. Detailed parametric design parameters for the CSRR are provided in Table II.

To further quantify the agreement between simulated and reference responses, we report the normalized root-mean-square error (*nrmse*), pearson correlation coefficient (*r*), and associated p-value (*p*) for both sensor types.

*D. Experimental validation on fetal bovine serum solution*

For experimental validation, the leaky-wave antenna and the CSRR-based resonator were fabricated, as illustrated in Figs. 5(a) and 5(b). A glucose-supplemented fetal bovine serum (FBS) solution was employed as the test sample to replicate the dielectric characteristics of human blood. Fig. 5(c) depicts the experimental setup used during testing, where the sensor was loaded with the FBS sample. To hold the fetal bovine serum (FBS) sample, a polypropylene (PP) container with dimensions of 71 mm × 37 mm × 17 mm (L × W × H) was positioned directly above the sensor. To assess the sensing capability, FBS-glucose solutions with different concentrations were prepared to mimic variations in physiological blood glucose levels.

Glucose concentrations ranging from 50 to 255 mg/dL were selected to enable detection of subtle variations within the typical physiological range of human blood glucose. To prepare the stock solution, 1150 mg of glucose was precisely weighed using an electronic balance and dissolved in 50 mL of deionized water, yielding a 2300 mg/dL glucose solution. Subsequently,

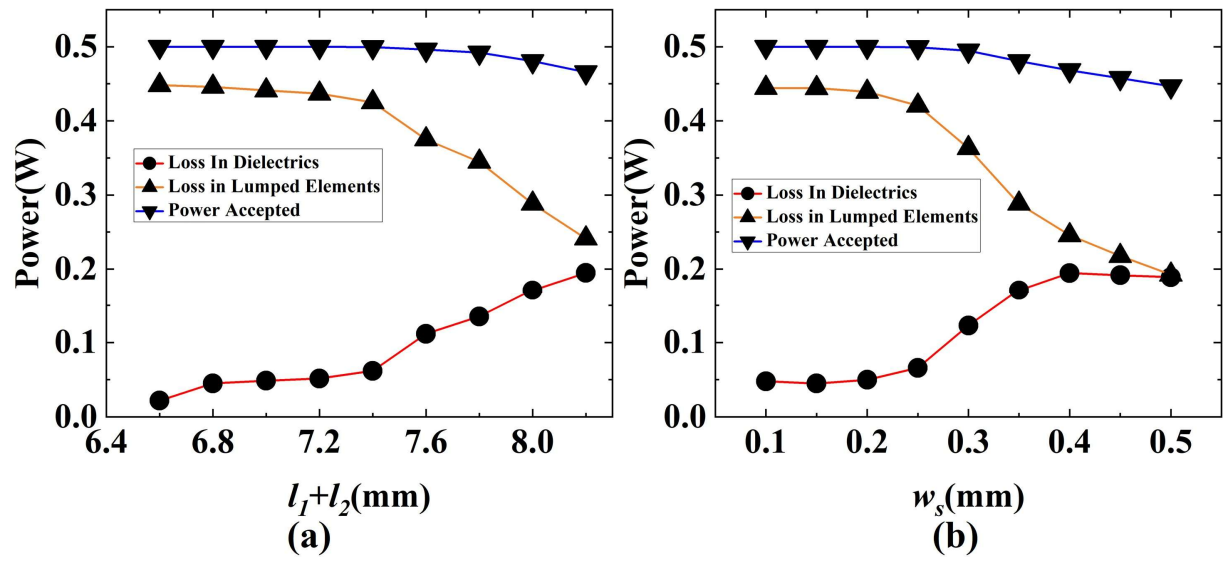

**Fig. 6.** Parametric optimization and phase sensitivity analysis of the leaky-wave antenna (LWA). (a) Variation of accepted power, dielectric loss, lumped loss, and phase with total slot length $l_1 + l_2$ and (b) slot width $w_s$.

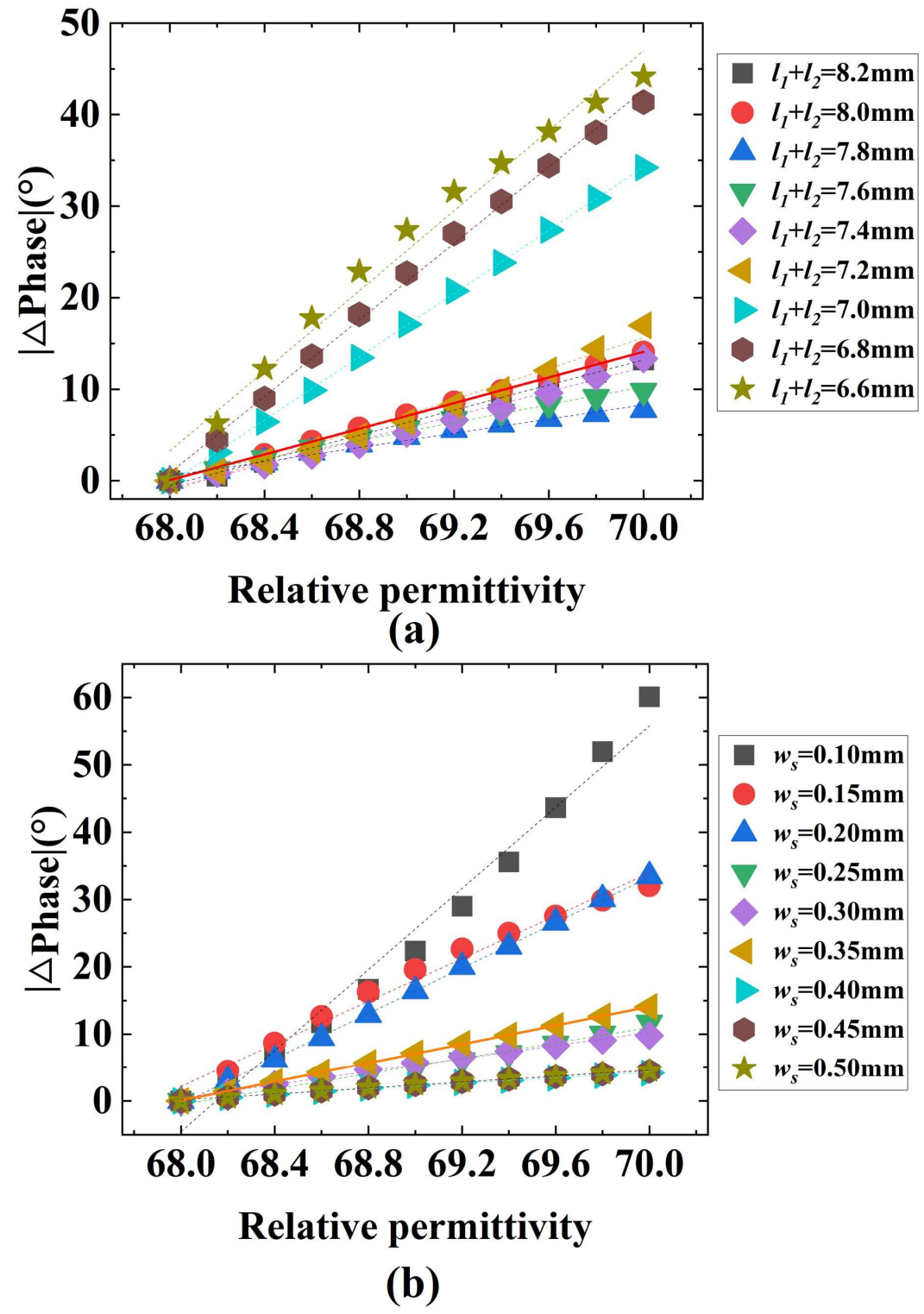

**Fig. 7.** Phase sensitivity of the LWA at different (a) $l_1 + l_2$ and (b) $w_s$ values across permittivity values from 68 to 70.

20 mL of fetal bovine serum (FBS) was measured using a syringe, and the glucose stock solution was incrementally added to create test samples with target glucose concentrations.

Prior to measurement, the vector network analyzer (VNA) was preheated for 30 minutes and calibrated to ensure measurement accuracy. Following calibration, the sensor was securely mounted on a foam base and connected to the VNA. The experimental setup for measuring fetal bovine serum (FBS) solutions with varying glucose concentrations is shown in Fig. 5(c).

A custom plastic container was placed above the sensor to minimize errors caused by mechanical disturbances during measurement. Initially, a base solution was prepared by dissolving 50 mg of anhydrous glucose powder in 100 mL of fetal bovine serum (FBS), yielding a starting concentration of 50 mg/dL. After thorough mixing, 20 mL of this base solution was transferred to the container, and the sensor's response to the initial concentration was recorded.

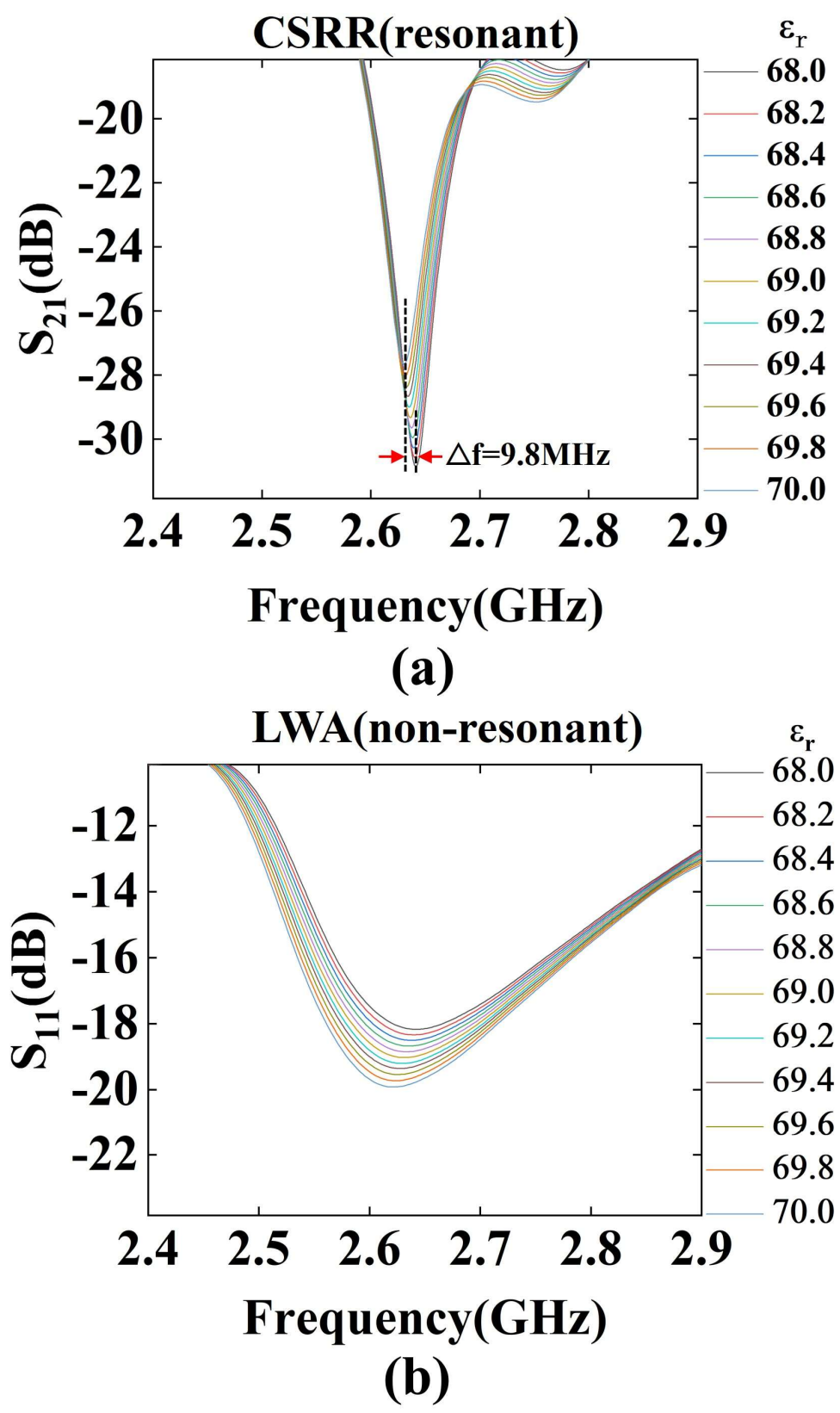

**Fig. 8.** Simulated S-parameters under loaded conditions showing amplitude variation with dielectric constant: (a) S21 of the CSRR and (b) S11 of the LWA at 2.63 GHz.

To introduce controlled, incremental changes in glucose concentration, a micropipette was used to add 0.2 mL of a 2300 mg/dL glucose solution to the base FBS sample. This approach ensured precise concentration adjustments and complete dissolution of the added glucose. The solution was then stirred thoroughly and allowed to stand for 10 minutes to ensure homogeneity before the vector network analyzer (VNA) recorded the amplitude and phase of the S-parameters.

This process was repeated ten times, resulting in the addition of a total of 2 mL of the concentrated glucose solution. The final glucose concentration in the mixed solution reached 255 mg/dL.

The dielectric sensitivity was quantified using frequency sensitivity ($S_{fre}$) and phase sensitivity ($S_{pha}$), defined as:

$$S_{fre} = \frac{\Delta f_r}{\Delta \varepsilon_r} = \frac{\left| f_{r(\varepsilon_r=70)} - f_{r(\varepsilon_r=68)} \right|}{2} \tag{5}$$

$$S_{pha} = \frac{\Delta S}{\Delta \varepsilon_r} = \frac{\left| S_{(\varepsilon_r=70)} - S_{(\varepsilon_r=68)} \right|}{2} \tag{6}$$

### *E. Experimental validation on human subjects*

Two healthy volunteers (2 males; aged 20–30 years) participated in multiple modified oral glucose tolerance tests (OGTTs), each involving ingestion of a 75 g glucose load. The ingestion period was standardized to 10–15 minutes to ensure consistency across participants. The study was approved by the ShanghaiTech University Scientific Research Ethics Review Committee (Approval No. Q2025-012), and written informed consent was obtained from all subjects. Eligible participants (aged 20–50 years, with no restrictions on race, sex, or ethnicity) were monitored in a controlled indoor environment (constant temperature and humidity) and remained seated throughout the procedure to minimize motion artifacts and sweat-induced ionic variations. To standardize epidermal hydration, the measurement site was sanitized with alcohol wipes. Ten minutes before glucose ingestion, sensors were affixed to the index finger (fully covering the sensing area) using wrap-around tape, as shown in Fig. 5(d), allowing continuous signal acquisition during the 60-minute OGTT. The S-parameters of the sensor were recorded every 5 minutes using a commercial vector network analyzer (Rohde & Schwarz ZNB40 Vector Network Analyzer) after standard calibration, with cables and sensors employing physical stabilization to prevent dynamic phase errors. This 5-minute sampling protocol using a laboratory-grade VNA was specifically adopted for early-stage proof-of-concept validation. Simultaneously, reference blood glucose levels were measured at an initial fasting baseline (t=0) and subsequently every 15 minutes post-ingestion using a standard finger-stick method with a commercial glucometer (Accu-Chek, Roche Ltd.) [14], [25]. To align with the 5-minute VNA measurements, intermediate glucose values were estimated via spline interpolation.

## III.RESULTS AND DISCUSSIONS

### *A. Optimization of LWA sensor*

To enhance the dielectric sensing performance of the leaky-wave antenna (LWA), we performed a structural optimization focusing on two critical parameters: the total slot length ($l_1 + l_2$) and the slot width ($w_s$). As illustrated in Fig. 6 and Fig. 7, varying these parameters affects the accepted power, dielectric loss, and phase response. Specifically, the configuration with $l_1 + l_2$= 8.2 mm and $w_s$= 0.35 mm was selected, as it provides a balanced trade-off between efficient power absorption, minimal internal loss, and stable phase variation.

The simulated S-parameter results for both sensor configurations are presented in Fig. 8. In order to illustrate high sensitivity, achieving a linear phase response is crucial for accurate and robust modeling. Fig. 9 shows the phase responses of the CSRR and LWA, respectively, across relative permittivity values from 68 to 70. The LWA exhibits a strong linear correlation between dielectric constant and phase shift, while the CSRR response exhibits nonlinear behavior under the tested configuration and operating conditions. At 2.63 GHz, linear fitting reveals a maximum phase deviation of only 0.120° for the LWA (Fig.9(b)) compared to 3.391° for the CSRR (Fig.9(a)). The LWA model further achieves a normalized root mean square error (*nrmse*) nearly 0.005 and a correlation coefficient (*r*) of 0.999, confirming high prediction accuracy and statistical significance ($p < 0.010$).

Furthermore, the reflection mode ($S_{11}$) modes of the optimized LWA exhibits strong linear correlations with dielectric constant changes (with achieves a normalized root mean square error (*nrmse*) nearly 0.004 and correlation coefficients $r$ = 0.999 and $p$ < 0.010). Specifically, the total phase variation of $S_{11}$ reaches approximately 15° across $\varepsilon_r$= 68–70.

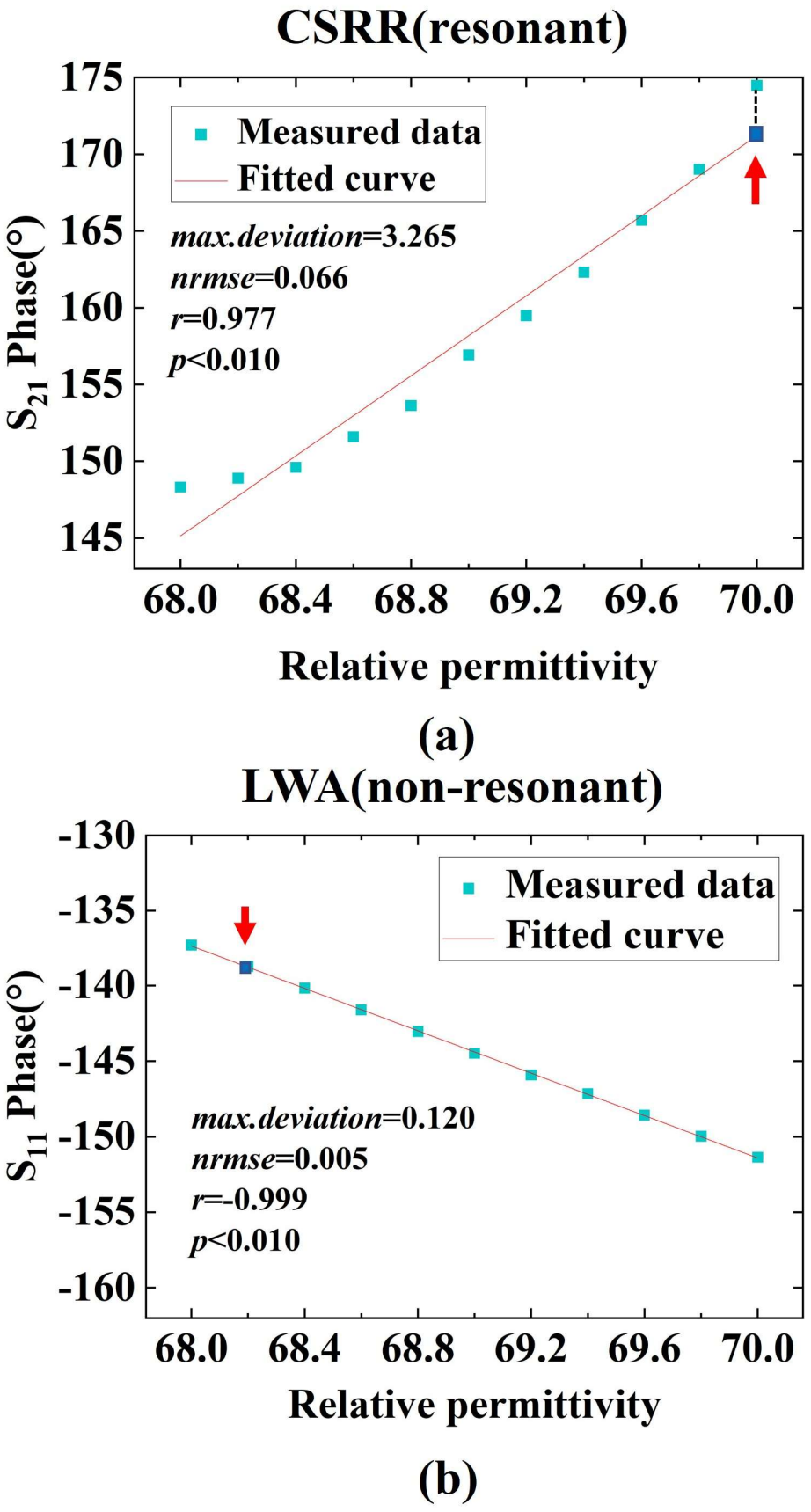

**Fig. 9.** Linear correlation between relative permittivity of the sample (a) $S_{21}$ phase of the CSRR, and (b) $S_{11}$ phase of the LWA at 2.63 GHz according to simulation results.

These results confirm that the optimized LWA not only provides higher dielectric sensitivity than the CSRR structure but also enables simplified and accurate modeling due to its highly linear phase response within the physiological range [23].

### *B. Resonant vs non-resonant sensors*

Compared to resonant antennas such as CSRR structures, the non-resonant leaky-wave antenna (LWA) demonstrates exhibits stable frequency behavior, offering an alternative sensing strategy against frequency drift and sample-induced resonance shifts. As shown in Fig. 8(a), the operating frequency of the CSRR sensor shifts significantly with changes in the sample's dielectric properties, often requiring high-performance vector network analyzers with high spectral resolution to accurately track the resonant peak. Simulation results show that the CSRR sensor exhibits a frequency sensitivity of 9.8 MHz. In contrast, as illustrated in Fig. 8(b), the LWA operates over a broad frequency range with minimal deviation, maintaining a stable operating point throughout the measurement. This frequency stability greatly simplifies signal acquisition and system implementation.

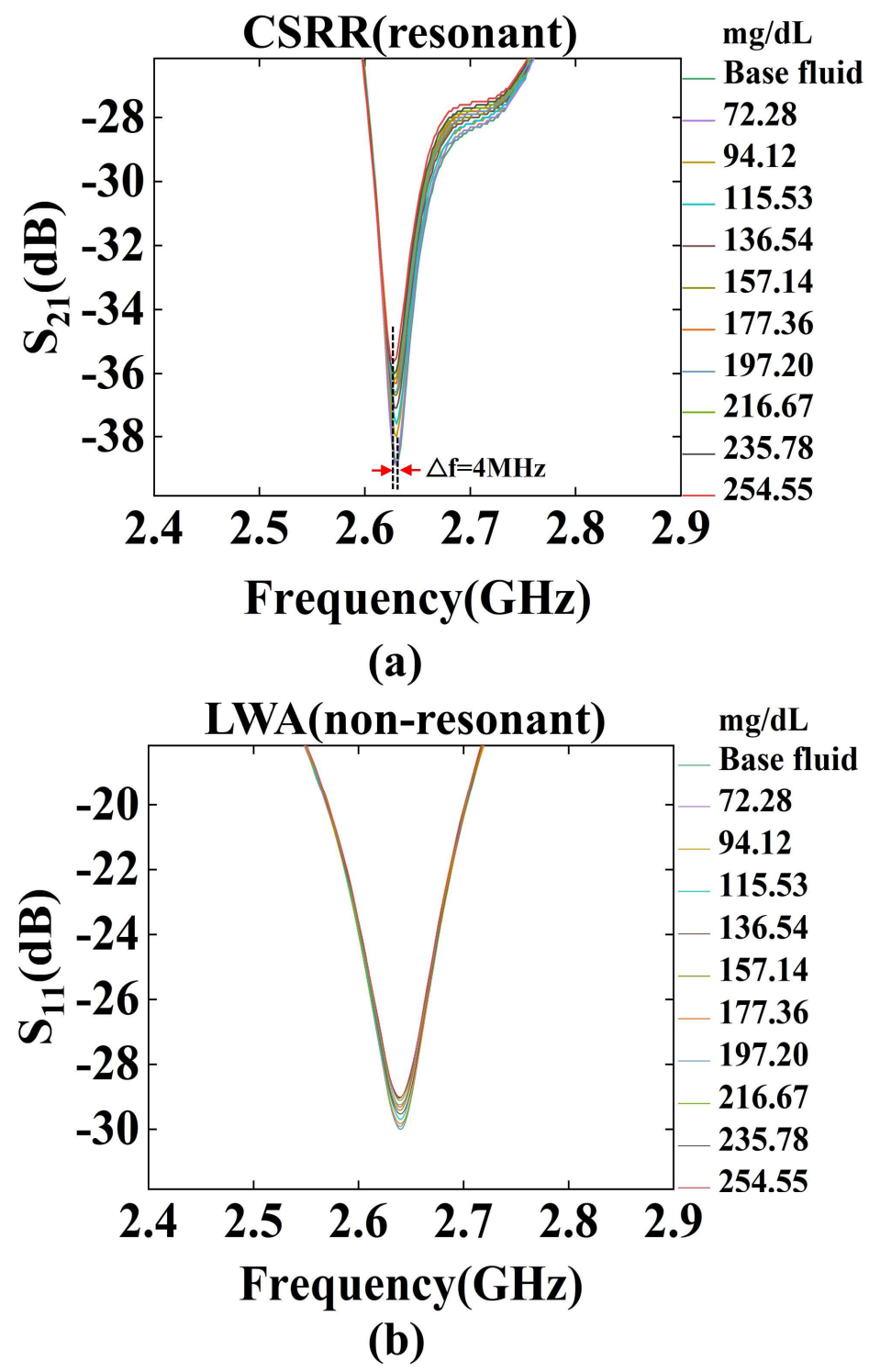

**Fig. 10.** (a) Measured transmission responses $S_{21}$ of the CSRR loaded with fetal bovine serum at different glucose concentrations. (b) Measured transmission responses $S_{11}$ of the LWA loaded with fetal bovine serum at different glucose concentrations.

This difference directly affects modeling strategies. Although the resonant frequency of CSRR is permittivity-dependent, its amplitude and phase at a fixed frequency show nonlinear behavior, which may introduce complexity in single-frequency modeling depending on the target range. By contrast, the LWA exhibits a clearly defined and approximately linear relationship between dielectric constant and phase (Fig. 9(b)), enabling accurate glucose estimation via simple linear models. This makes LWA more robust and better suited for real-time, low-complexity sensing.

### *C. Experimental validation on fetal bovine serum solution*

As shown in Fig. 10(a), experimental results for the CSRR sensor indicate that its resonance frequency shifts in response to changes in the sample's dielectric constant. The CSRR exhibits a nominal resonance at approximately 2.63 GHz. As the glucose concentration in fetal bovine serum increases from 50 mg/dL to 255 mg/dL, the resonance frequency shifts from 2.63 GHz to approximately 2.626 GHz, corresponding to a total frequency shift of approximately 4 MHz. The resulting frequency sensitivity is calculated to be $1.951 \times 10^{-2}$ MHz/(mg/dL). Phase fitting analysis further reveals a maximum deviation of 0.606° from the ideal response for the LWA, in contrast to 2.584° for the CSRR.

As shown in Fig. 10(b), the LWA sensor maintains a stable operating frequency near 2.63 GHz across varying glucose concentrations. Additionally, it shows that the $S_{11}$ amplitude at 2.63 GHz decreases from −30.20 dB to −29.12 dB as the glucose concentration increases from 50 mg/dL to 255 mg/dL, resulting in an amplitude change of approximately 1.08 dB. This corresponds to an amplitude sensitivity of $5.268 \times 10^{-3}$ dB/(mg/dL). Additionally, the phase of the $S_{11}$ parameter increases from −83.92° to −75.41°, yielding a phase shift of 8.51° and a corresponding phase sensitivity of 0.0415°/(mg/dL). This phase sensitivity outperforms other state-of-the-art microwave-based glucose sensors, as summarized in Table III.

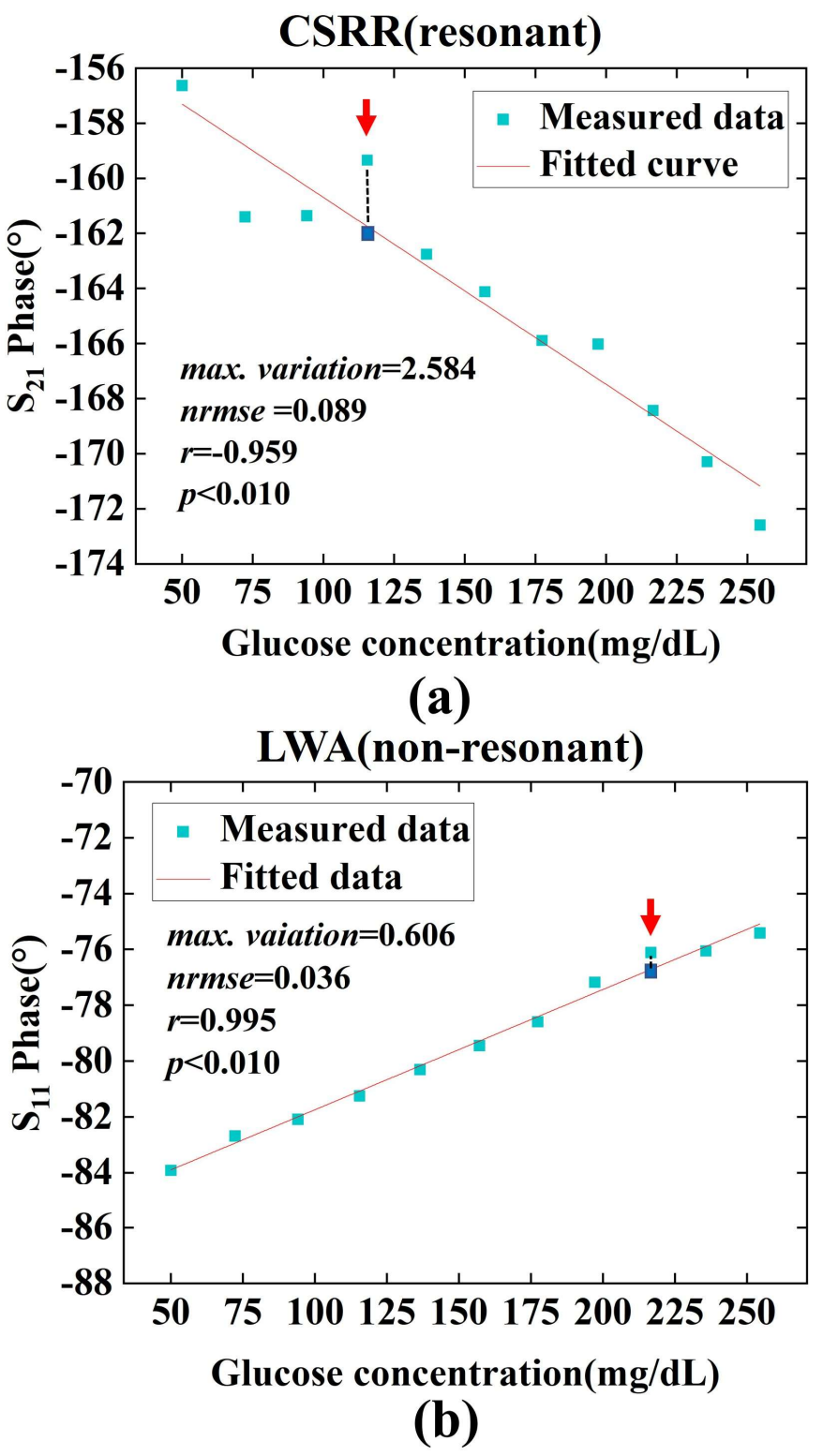

**Fig. 11.** (a) Linear correlation between relative permittivity of the sample and $S_{21}$ phase of the CSRR at 2.63 GHz according to experimental results. (b) Linear correlation between relative permittivity of the sample and $S_{11}$ phase of the LWA at 2.63 GHz according to experimental results.

As shown in Fig. 11, the leaky-wave antenna exhibits a strong linear correlation between phase response and glucose concentration at 2.63 GHz. The CSRR demonstrates a less linear phase response in this specific glucose-relevant permittivity range.

### D. Experimental validation on human subjects

To further validate the proposed concept, we performed in vivo experiments on healthy human subjects in which the phase of the coupling coefficient $S_{11}$ was measured using a vector network analyzer. The measured $S_{11}$ phase was compared with blood glucose concentrations obtained from a reference glucose meter, with results shown in Fig. 12. It is noteworthy that the two trials yielded different average $S_{11}$ phase values and glucose levels, reflecting the high sensitivity of these parameters to individual physiological conditions and timing of the measurements. For Trial 1 and Trial 2, the correlation coefficients between the blood glucose concentrations and the measured $S_{11}$ phase were 0.939 and 0.906, respectively.

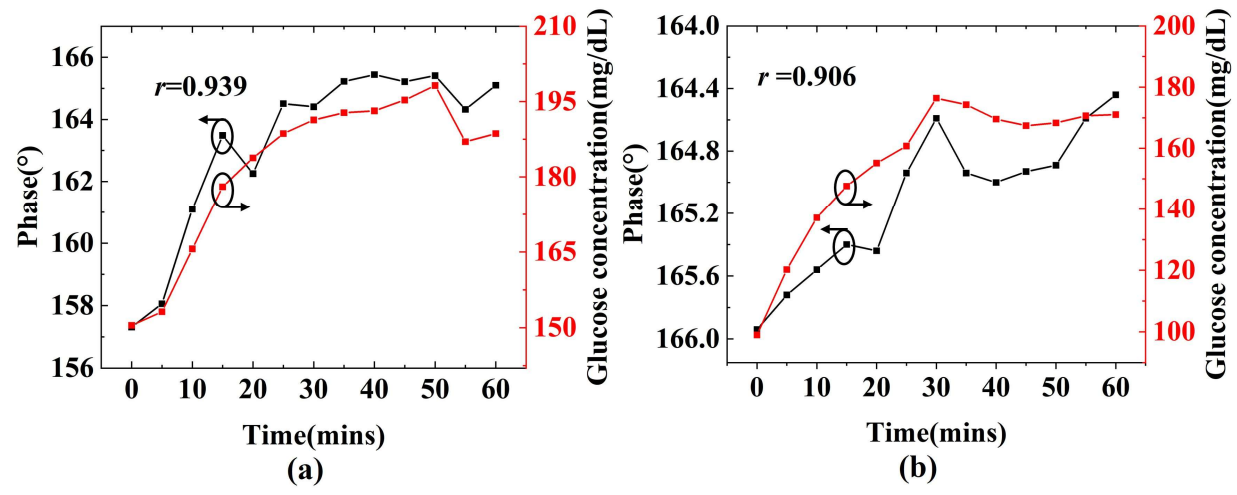

**Fig. 12.** Measured temporal evolution of $S_{11}$ phase in comparison to glucose levels. (a) Trial 1. (b) Trial 2.

TABLE III
COMPARISON BETWEEN STATE-OF-THE-ART MICROWAVE SENSORS USED FOR GLUCOSE DETECTION

| Ref | Concentration (mg/dL) | Operating frequency (GHz) | Sensing parameter | Sensitivity(per mg/dL) |
|---|---|---|---|---|
| [26] | 0-70000 | 6.5 | $\Delta f_r(S_{21})$ | 3.13E-05 (MHz ) |
| [27] | 500-20000 | 35.5 | $\Delta f_r(S_{21})$ | 4.00E-05(MHz) |
| [28] | 504-3531 | 1.4 | $\Delta f_r(S_{21})$ | 1.83E-04(MHz) |
| [29] | 500-2500 | 1.91 | $\Delta f_r(S_{21})$ | 4.00E-04(MHz) |
| [30] | 0-1000 | 3.1-3.8 | $\Delta f_r(S_{11})$ | 1.40E-02(MHz) |
| [31] | 50-40000 | 14-16 | Phase($S_{11}$) | 2.40E-02(°) |
| This work | 55-255 | 2.63(non-resonant) | Phase($S_{11}$) | 4.15E-02(°) |

While absolute phase values differed between subjects due to physiological baselines, the sensor consistently tracked glycemic changes for each individual. Given the preliminary nature of this engineering proof-of-concept study and the severely limited sample size (N=2), comprehensive clinical accuracy metrics were not evaluated. Instead, the fundamental efficacy of the sensor was assessed based on its correlational ability to trend with physiological changes. As such, larger-scale clinical studies with substantially expanded subject numbers and measurement repetitions will be required before any clinically meaningful performance evaluation can be made. While these strong temporal correlations demonstrate the underlying technical feasibility of the LWA sensing mechanism under strictly controlled measurement conditions, transitioning to unconstrained environments remains a challenge. Consistent with established practices in recent proof-of-concept studies [14], [32], our future work will focus on conducting dynamic stress tests and integrating auxiliary multi-modal sensors (such as temperature, pressure, and motion sensors) to enable active compensation. These future steps are essential to establish a comprehensive uncertainty budget against physiological and environmental confounding factors prior to clinical translation. This includes a systematic evaluation of repeatability, repositioning error, contact-pressure variation, sample variability, and long-term drift. Furthermore, a systematic characterization of hardware-related non-idealities, such as termination mismatch, will be conducted to pave the way for future practical wearable implementation.

## IV. Conclusion

This paper has introduced and experimentally validated a novel non-resonant leaky-wave antenna that addresses stability considerations associated with resonance-tracking microwave sensors by offering a propagation-based alternative. The proposed sensor achieved a phase sensitivity of 0.0415 °/(mg/dL) in vitro and demonstrated strong temporal correlations with reference glucose levels in preliminary human proof-of-concept trials. The inherent robustness of the traveling-wave mechanism, combined with these promising results, establishes this approach as a significant proof-of-concept step toward practical, low-cost, and wearable continuous glucose-monitoring systems, which will require future integration with miniaturized RF hardware.

## References

[1] P. K. Crane et al., “Glucose levels and risk of dementia,” N. Engl. J. Med., vol. 369, no. 6, pp. 540–548, Aug. 2013, doi: 10.1056/NEJMoa1215740.

[2] American Diabetes Association, “7. Diabetes technology: Standards of Medical Care in Diabetes—2020,” Diabetes Care, vol. 43, no. Suppl. 1, pp. S77–S88, Jan. 2020, doi: 10.2337/dc20-S007.

[3] D. S. Feig et al., “Continuous glucose monitoring in pregnant women with type 1 diabetes (CONCEPTT): a multicentre international randomised controlled trial,” The Lancet, vol. 390, no. 10110, pp. 2347–2359, Nov. 2017, doi: 10.1016/S0140-6736(17)32400-5.

[4] L. Tang, S. J. Chang, C.-J. Chen, and J.-T. Liu, “Non-Invasive Blood Glucose Monitoring Technology: A Review,” Sensors, vol. 20, no. 23, p. 6925, Dec. 2020, doi: https://doi.org/10.3390/s20236925.

[5] J. Blauert and Asimina Kiourti, “Bio-Matched Horn: A Novel 1–9 GHz On-Body Antenna for Low-Loss Biomedical Telemetry With Implants,” IEEE Transactions on Antennas and Propagation, vol. 67, no. 8, pp. 5054–5062, Aug. 2019, doi: https://doi.org/10.1109/tap.2018.2889159.

[6] X. Y. Liu, Z. T. Wu, Y. Fan, and E. M. Tentzeris, “A miniaturized CSRR loaded Wide-Beamwidth circularly polarized implantable antenna for subcutaneous Real-Time glucose monitoring,” IEEE Antennas and Wireless Propagation Letters, vol. 16, pp. 577–580, Jul. 2016, doi: 10.1109/lawp.2016.2590477.

[7] H. Xi and Y. Gong, “Optical non-invasive glucose monitoring using aqueous humor: A review,” Sensors, vol. 25, Art. no. 4236, Jul. 2025, doi: 10.3390/s25134236.

[8] Q. Zhai et al., “Enokitake mushroom-like standing gold nanowires toward wearable noninvasive bimodal glucose and strain sensing,” ACS Appl. Mater. Interfaces, vol. 11, no. 10, pp. 9724–9729, 2019.

[9] Y. Chen et al., “Skin-like biosensor system via electrochemical channels for noninvasive blood glucose monitoring,” Sci. Adv., vol. 3, no. 12, Art. no. e1701629, Dec. 2017, doi: 10.1126/sciadv.1701629.

[10] A. Kandwal, L. W. Liu, M. J. Deen, R. Jasrotia, B. K. Kanaujia, and Z. Nie, “Electromagnetic wave sensors for noninvasive blood glucose monitoring: Review and recent developments,” IEEE Trans. Instrum. Meas., vol. 72, pp. 1–15, 2023, doi: 10.1109/TIM.2023.3327466.

[11] M. C. Cebedio, L. A. Rabioglio, I. E. Gelosi, R. A. Ribas, A. J. Uriz, and J. C. Moreira, “Analysis and design of a microwave coplanar sensor for non-invasive blood glucose measurements,” IEEE Sensors J., vol. 20, no. 18, pp. 10572–10581, Sep. 2020.

[12] R. Zhang et al., “Noninvasive electromagnetic wave sensing of glucose,” Sensors, vol. 19, no. 5, Art. no. 1151, Mar. 2019, doi: 10.3390/s19051151.

[13] S. Kiani, P. Rezaei, and M. Fakhr, “Dual-frequency microwave resonant sensor to detect noninvasive glucose-level changes through the fingertip,” IEEE Trans. Instrum. Meas., vol. 70, pp. 1-8, 2021.

[14] J. Hanna et al., “Noninvasive, wearable, and tunable electromagnetic multisensing system for continuous glucose monitoring, mimicking vasculature anatomy,” Sci. Adv., vol. 6, no. 24, Jun. 2020, Art. no. eaba5320.

[15] S. Wang, W. Wang, and Y. Zheng, “Reconfigurable multimode microwave sensor with resonance and transmission sensing capabilities for noninvasive glucose monitoring,” IEEE Trans. Microwave Theory Techn., vol. 72, no. 5, pp. 3102–3117, May 2024, doi: 10.1109/TMTT.2023.3324432.

[16] N. K. Tiwari, A. K. Jha, S. P. Singh, and M. Jaleel Akhtar, “Estimation of broadband complex permeability using SIW cavity-based multimodal approach,” IEEE Trans. Instrum. Meas., vol. 69, no. 9, pp. 6571–6581, Sep. 2020, doi: 10.1109/TIM.2020.2969590.

[17] M. Bteich et al., “A non-invasive flexible glucose monitoring sensor using a broadband reject filter,” IEEE J. Electromagn. RF Microw. Med. Biol., vol. 5, no. 2, pp. 139–147, Jun. 2021, doi: 10.1109/JERM.2020.3023053.

[18] J. B. Pendry, A. J. Holden, D. J. Robbins, and W. J. Stewart, “Magnetism from conductors and enhanced nonlinear phenomena,” IEEE Trans. Microwave Theory Techn., vol. 47, no. 11, pp. 2075–2084, Nov. 1999, doi: 10.1109/22.798002.

[19] F. Falcone, T. Lopetegi, J. D. Baena, R. Marqués, F. Martín, and M. Sorolla, “Effective negative-ε stopband microstrip lines based on complementary split ring resonators,” IEEE Microw. Wireless Compon. Lett., vol. 14, no. 6, pp. 280–282, Jun. 2004, doi: 10.1109/LMWC.2004.828029.

[20] H. Xiao and Y. Gao, “Novel complementary split-ring resonator sensors for high-sensitivity permittivity characterization,” in Proc. 2023 IEEE 11th Asia-Pacific Conf. Antennas Propag. (APCAP), 2023, doi: 10.1109/APCAP59480.2023.10470248.

[21] O. Losito and V. Dimiccoli, “Travelling planar wave antenna for wireless communications,” in Wireless Communications and Networks - Recent Advances, A. Eksim, Ed., InTech, 2012. doi: 10.5772/35914.

[22] G.-B. Wu, Q.-L. Zhang, K. F. Chan, B.-J. Chen, and C. H. Chan, “Amplitude-modulated leaky-wave antennas,” IEEE Trans. Antennas Propagat., vol. 69, no. 7, pp. 3664–3676, Jul. 2021, doi: 10.1109/TAP.2020.3044673.

[23] T. Karacolak, E. C. Moreland, and E. Topsakal, “Cole–cole model for glucose-dependent dielectric properties of blood plasma for continuous glucose monitoring,” Microw. Opt. Technol. Lett., vol. 55, no. 5, pp. 1160–1164, May 2013, doi: 10.1002/mop.27515.

[24] K. Abdesselam et al., “A non-invasive honey-cell CSRR glucose sensor: Design considerations and modelling,” IRBM, vol. 44, no. 1, Art. no. 100713, Feb. 2023, doi: 10.1016/j.irbm.2022.04.002.

[25] H. M. Farhat et al., “Wireless continuous glucose monitoring using V-band circularly polarized directive antenna arrays in a transmit/receive system,” IEEE Trans. Antennas Propagat., vol. 72, no. 9, pp. 7005–7016, Sep. 2024, doi: 10.1109/TAP.2024.3436578.

[26] A. K. Jha et al., “Broadband wireless sensing system for non-invasive testing of biological samples,” IEEE J. Emerg. Sel. Topics Circuits Syst., vol. 8, no. 2, pp. 251–259, Jun. 2018, doi: 10.1109/JETCAS.2018.2829205.

[27] A. I. Gubin et al., “Whispering-gallery-mode resonator technique with microfluidic channel for permittivity measurement of liquids,” IEEE Trans. Microwave Theory Techn., vol. 63, no. 6, pp. 2003–2009, Jun. 2015, doi: 10.1109/TMTT.2015.2423289.

[28] H. Choi et al., “Design and in vitro interference test of microwave noninvasive blood glucose monitoring sensor,” IEEE Trans. Microwave Theory Techn., vol. 63, no. 10, pp. 3016–3025, Oct. 2015, doi: 10.1109/TMTT.2015.2472019.

[29] G. Gennarelli, S. Romeo, M. R. Scarfi, and F. Soldovieri, “A microwave resonant sensor for concentration measurements of liquid solutions,” IEEE Sensors J., vol. 13, no. 5, pp. 1857–1864, May 2013.

[30] V. Turgul and I. Kale, “Simulating the effects of skin thickness and fingerprints to highlight problems with non-invasive RF blood glucose sensing from fingertips,” IEEE Sensors J., vol. 17, no. 22, pp. 7553–7560, 2017.

[31] M. Hofmann, G. Fischer, R. Weigel, and D. Kissinger, “Microwave-based noninvasive concentration measurements for biomedical applications,” IEEE Trans. Microw. Theory Techn., vol. 61, no. 5, pp. 2195–2204, May 2013.

[32] M. A. Nguyen, F. Bien, and G. Byun, "Design of an Out-Folded Patch Antenna With a Zeroth-Order Resonance for Non-Invasive Continuous Glucose Monitoring," IEEE Trans. Antennas Propag., vol. 70, no. 10, pp. 8932-8940, Oct. 2022.